\def\year{2015}
\documentclass[letterpaper]{article}
\usepackage{aaai}
\usepackage{times}
\usepackage{helvet}
\usepackage{courier}
\usepackage{graphicx}
\usepackage{color,soul}
\usepackage{multicol}
\begin{document}

\title{Towards Detecting Rumours in Social Media}

\author{Arkaitz Zubiaga$^1$, Maria Liakata$^1$, Rob Procter$^1$, Kalina Bontcheva$^2$, Peter Tolmie$^1$\\
$^1$University of Warwick, UK\\
$^2$University of Sheffield, UK\\
\{a.zubiaga,m.liakata,rob.procter\}@warwick.ac.uk, k.bontcheva@sheffield.ac.uk, peter.tolmie@gmail.com
}
\maketitle
\begin{abstract}
\begin{quote}
The spread of false rumours during emergencies can jeopardise the well-being of citizens as they are monitoring the stream of news from social media to stay abreast of the latest updates. In this paper, we describe the methodology we have developed within the PHEME project for the collection and sampling of conversational threads, as well as the tool we have developed to facilitate the annotation of these threads so as to identify rumourous ones. We describe the annotation task conducted on threads collected during the 2014 Ferguson unrest and we present and analyse our findings. Our results show that we can collect effectively social media rumours and identify multiple rumours associated with a range of stories that would have been hard to identify by relying on existing techniques that need manual input of rumour-specific keywords.
\end{quote}
\end{abstract}

\section{Introduction}

While the spread of inaccurate or questionable information has always been a concern, the emergence of the Internet and social media has exacerbated the problem by facilitating the spread of such information to large communities of users \cite{koohang2003misinformation}. This is especially the case in emergency situations, where the spread of a false rumour can have dangerous consequences. For instance, in a situation where a hurricane is hitting a region, or a terrorist attack occurs in a city, access to accurate information is crucial for finding out how to stay safe and for maximising citizens' well-being. This is even more important in cases where users tend to pass on false information more often than real facts, as occurred with Hurricane Sandy in 2012 \cite{zubiaga2014tweet}. Hence, identifying rumours within a social media stream can be of great help for the development of tools that prevent the spread of inaccurate information.

The first step in a study of social media rumours is the identification of an appropriate dataset that includes a diverse set of stories. To-date, related work has relied on picking out rumours through manual identification of well-known viral stories \cite{qazvinian2011rumor,castillo2013predicting,procter2013reading}, in some cases focusing only on false rumours \cite{starbird2014rumors}. In these cases, the authors defined a specific set of keywords that were known to be related to rumourous stories and harvested the tweets containing those keywords. However, the process of carefully defining what rumours are, as well as setting forth a sound methodology to identify rumours as an event unfolds, which would enable broader and deeper analysis of this phenomenon, remains unstudied. Our goal within the PHEME project\footnote{http://www.pheme.eu} is therefore to look closely at how rumours emerge in social media, how they are discussed and how their truthfulness is evaluated. In order to create social media rumour datasets, we propose an alternative way of manually annotating rumours by reading through the timeline of tweets related to an event and selecting stories that meet the characteristics of a rumour. This will enable not only the identification of a rich set of rumours, but also the collection of non-rumourous stories. The creation of both rumour and non-rumour datasets will allow us to train machine learning classifiers to assist with the identification of rumours in new events, by distinguishing the characteristics of threads that spark conversation from rumour-bearing ones. 

In this paper, we introduce a novel methodology to create a dataset of rumours and non-rumours posted in social media as an event unfolds. This methodology consists of three main steps: (i) collection of (source) tweets posted during an emergency situation, sampling in such a way that it is manageable for human assessment, while generating a good number of rumourous tweets from multiple stories, (ii) collection of conversations associated with each of the source tweets, which includes a set of replies discussing the source tweet, and (iii) collection of human annotations on the tweets sampled. We provide a definition of a rumour which informs the annotation process. Our definition draws on definitions from different sources, including dictionaries and related research. We define and test this methodology for tweets collected during the 2014 Ferguson unrest in the United States and present, analyse and discuss the outcome of the annotation task. We conclude the paper by discussing the effectiveness of the methodology, its application to the context of cities, and outline ongoing and future work analysing the evolution of and discussion around rumours in social media.

\section{Background}

While there is a substantial amount of research around rumours in a variety of fields, ranging from psychological studies \cite{rosnow2005rumor} to computational analyses \cite{qazvinian2011rumor}, defining and differentiating them from similar phenomena remains an active topic of discussion. Some researchers have attempted to provide a solid definition and characterisation of rumours so as to address the lack of common understanding around the specific categorisation of what is or is not a rumour. \cite{difonzo2007rumor} emphasise the need to differentiate rumours from similar phenomena such as gossip and urban legends. They define rumours as \textit{``unverified and instrumentally relevant information statements in circulation that arise in contexts of ambiguity, danger or potential threat and that function to help people make sense and manage risk''}. This definition also ties in well with that given by the Oxford English Dictionary (OED): \textit{``A currently circulating story or report of uncertain or doubtful truth''}\footnote{http://www.oxforddictionaries.com/definition/english/rumour}. Moreover, \cite{guerin2006analyzing} provide a detailed characterisation of rumours, highlighting the following points about a rumour: (i) it is of personal consequence and interest to listeners, (ii) the truth behind it is difficult to verify, (iii) it gains attention with horror or scandal, and (iv) it has to be new or novel.

There is a growing body of research on the analysis of rumours in the context of social media. Some researchers have looked at how social media users support or deny rumours in breaking news situations but their results are, as yet, inconclusive. In some cases it has been suggested that Twitter does well in debunking inaccurate information thanks to self-correcting properties of crowdsourcing as users share opinions, conjectures and evidence. For example, \cite{castillo2013predicting} found that the ratio between tweets supporting and debunking false rumours was 1:1 (one supporting tweet per debunking tweet) in the case of a 2010 earthquake in Chile. Procter et al. \cite{procter2013reading} came to similar conclusions in their analysis of false rumours during the 2011 riots in England, but they noted that any self-correction can be slow to take effect. In contrast, in their study of the 2013 Boston Marathon bombings, \cite{starbird2014rumors} found that Twitter users did not do so well in telling the truth from hoaxes. Examining three different rumours, they found the equivalent ratio to be 44:1, 18:1 and 5:1 in favour of tweets supporting false rumours.

These results provide evidence that Twitter’s self- correction mechanism cannot be relied upon in all circumstances, and suggest the need for more research that will help people to judge the veracity of rumours more quickly and reliably. This is the primary goal of the PHEME project and one of the first steps has been to define a methodology for selecting rumours associated with an event.  

In terms of rumour analysis, we are particularly interested in looking in detail at the conversational features of social media \cite{meredith2013}, so once we identify a tweet that introduces a rumour (i.e. the \textit{source} tweet), we then collect all tweets having a \textit{reply} relationship with the source tweet, to create a unit of tweets that we call a \textit{thread}.

\section{Annotation of Rumours}

We begin by expanding on the OED's definition with additional descriptions from rumour-related research, to provide a definition of rumour that is richer and, we argue, more appropriate for our purposes. We formally define a rumour as a \textit{circulating story of questionable veracity, which is apparently credible but hard to verify, and produces sufficient skepticism and/or anxiety}.

Previous work on annotation of rumourous stories from tweets \cite{qazvinian2011rumor,procter2013reading} has relied on the identification a priori of these stories -- i.e., by looking at media reports that summarise and debunk some of the rumours -- to define a set of relevant keywords for each rumour, and then filter tweets associated with those keywords. While this approach enables collection of a good number of tweets for each rumour, it does not guarantee the collection of a diverse set of stories associated with an event. Instead, we define keywords that broadly refer to an ongoing event, which is not a rumour itself but is expected to spark rumours. Having obtained collections of events, our work focuses on visualising the timeline of tweets associated with an event, to enable identification of rumourous content for a set of stories that is not necessarily known a priori, and that is therefore expected to generate a more diverse such set. For instance, \cite{starbird2014rumors} studied rumours from the 2013 Boston Marathon bombings by manually picking three well-known rumourous stories: (i) a girl was killed while running in the marathon, (ii) navy seals or craft security or blackwater agents as perpetrators, and (iii) the crowd misidentifies Sunil Tripathi as a bomber. While these three stories were widely discussed because of their popularity, and might provide a suitable scenario for certain studies, we are interested in identifying a broader set of rumourous stories in social media. Hence, we set out to enlist the help of experienced practitioners (i.e., journalists) to read through a timeline of tweets to identify rumours.

\subsection{Annotation Task}

In this annotation task, the human assessor reads through a timeline of tweets to determine which of these are associated with rumours. Without necessarily having prior knowledge of the rumours associated with a given event, we expect that this approach will let us discover new stories. To facilitate the task, we had to deal with two major issues: (i) the number of tweets tends to be large for any given event, and (ii) a tweet does not always provide enough context to be able to determine whether it is referring to a rumour.

\begin{figure}[htb]
 \centering
 \includegraphics[width=0.45\textwidth]{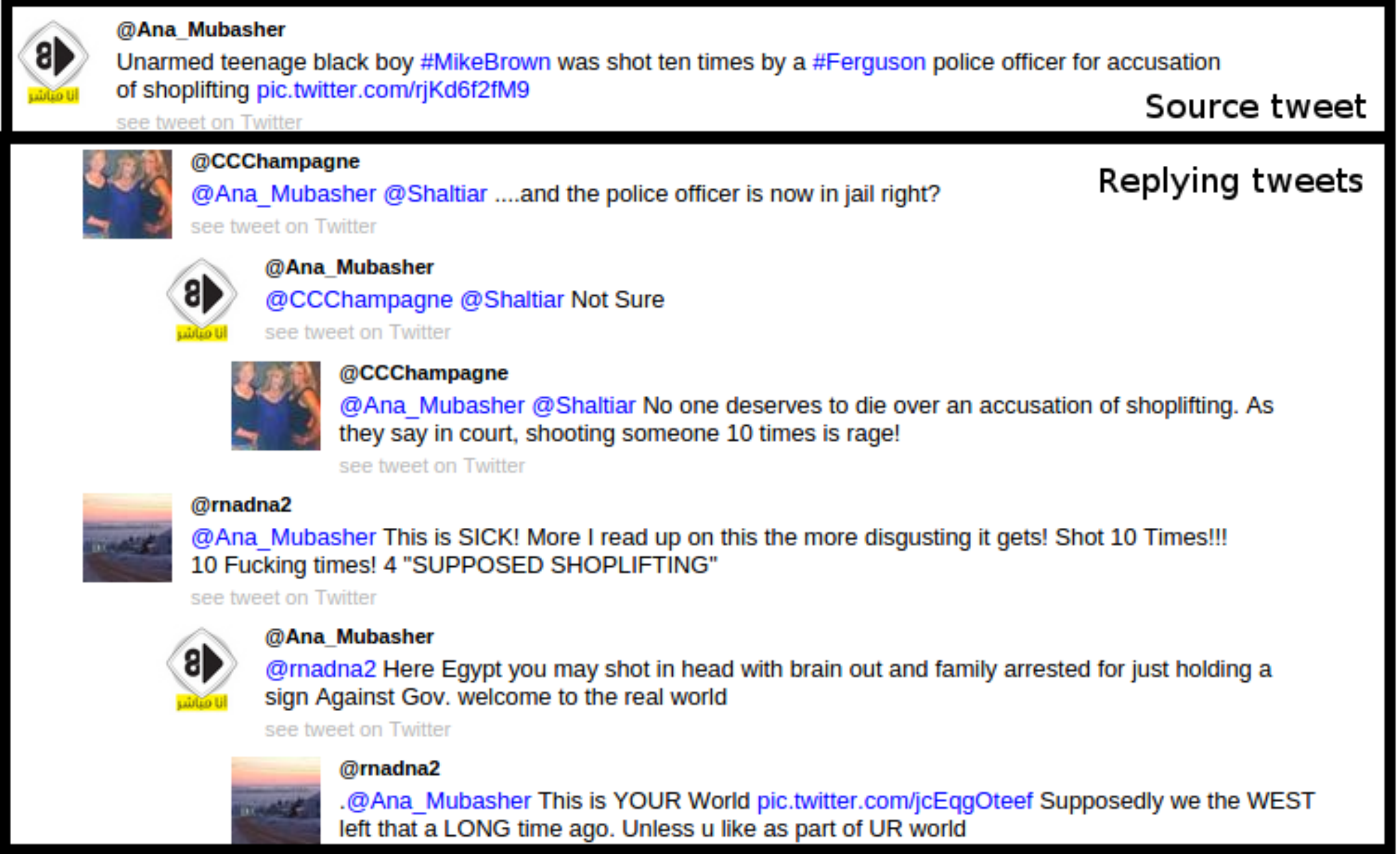}
 \caption{Example of a conversation sparked by a rumourous tweet, with a source tweet, and several tweets replying to it}
 \label{fig:source-and-replies}
\end{figure}

To alleviate the task and address (i) we reduced the number of tweets to be annotated, by employing a sampling technique that favours the presence of rumours, yet yields a set of tweets representative of the timeline of stories associated with the event in question. To achieve this, we rely on the characteristics of rumours to sample the data. By definition, a rumour has to generate significant interest within a community of users, which can be straightforwardly measured on Twitter by the number of times a tweet is retweeted. A tweet might introduce questionable information even without being shared massively, but it will not become a rumour until it is spread and further discussed by many. Hence, based on this assumption, we sample the tweets that exceed a specific number of retweets.

To enrich the inherently limited context of a tweet and address (ii) we look at tweets replying to it. While a tweet might not always help determine whether the underlying story was rumourous at the time of posting, the replies from others in case of a discussion can help provide clarity. Thus, we also collect tweets that reply to source tweets, as we will describe later. This allows us to have threads composed of a source tweet, which provides the starting point of a conversation, and a set of tweets which reply to that source tweet. Figure \ref{fig:source-and-replies} shows an example of a thread. We define each of these threads as the unit of the annotation task. The human annotator has to then look at the source tweet of the thread to determine if it is a rumour, and can optionally look at the conversation it sparked for more context.

\section{Rumour Annotation Tool}

\begin{figure*}[hbt]
 \centering
 \includegraphics[width=0.95\textwidth]{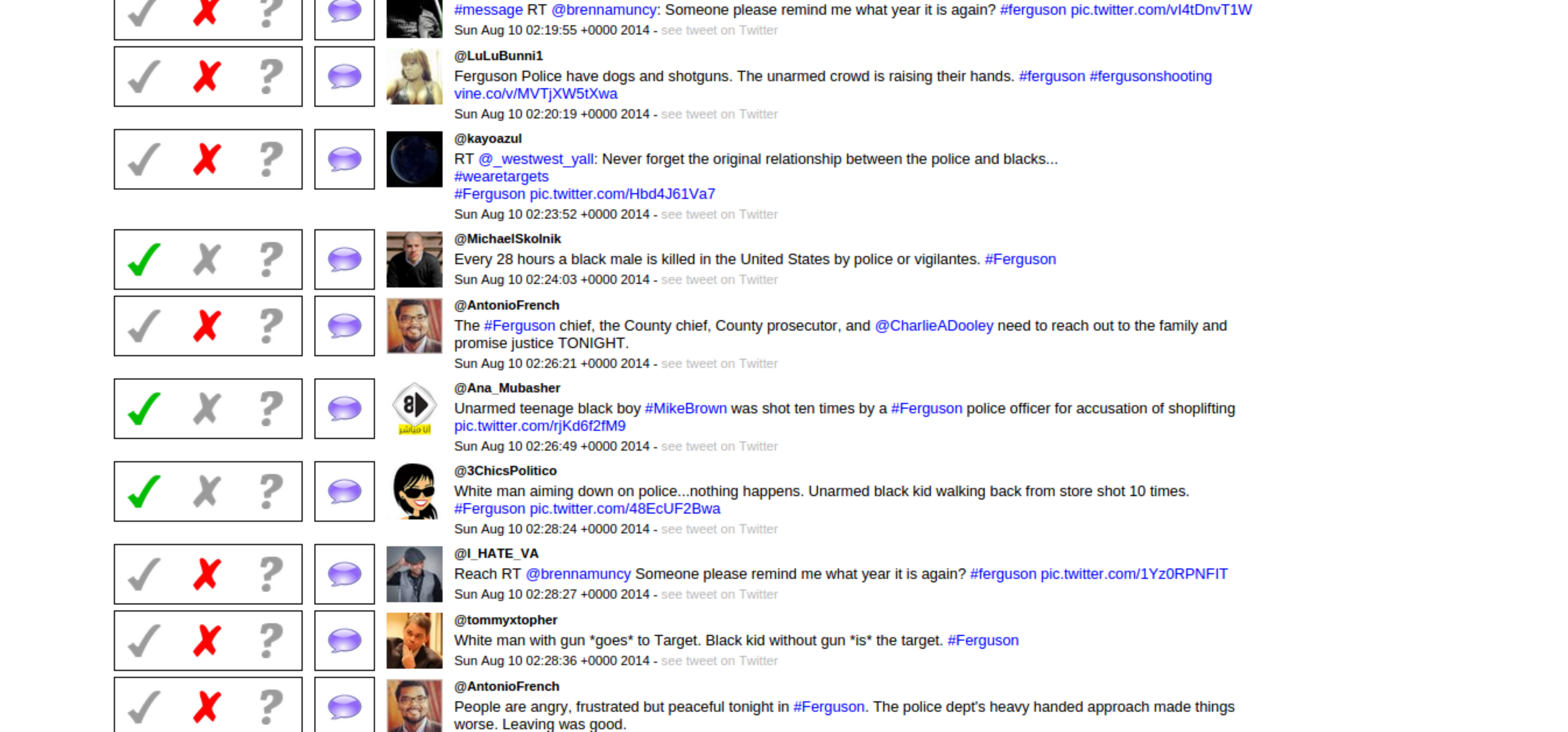}
 \caption{Rumour annotation tool, with tweets about the Ferguson unrest on the 10th of August}
 \label{fig:replies-retweets-sandy}
\end{figure*}

To facilitate the annotation task, we developed a tool that visualises the timeline of tweets associated with an event. The purpose of the tool is to enable annotators to read through the tweets and annotate them as being rumours or non-rumours. Annotators record their selections by clicking on the appropriate icon next to each source tweet (green tick for a rumour, a red cross for a non-rumour, or an orange question mark). Each source tweet is also accompanied by a bubble icon, which the annotator can click on to visualise the conversation sparked by a source tweet.

When the annotator marks a tweet as non-rumourous, the task for that tweet ends there. However, when they mark it as a rumour, the tool asks the annotator to specify the story associated with the rumour corresponding to that source tweet. Assigning a story to a rumour means that they categorise the rumourous tweet as being part of that story; a story is identified by a label that describes it. This way, annotators can group together tweets about the same rumour, and provide a descriptive label denoting what the story is about. This will allow us to study rumourous conversations separately, as well as examine them in the context of other conversations within the same story. In order to analyse the time taken to annotate each of the threads and assess the cost of the task, we save the timestamp every time the annotator makes a selection of rumour or non-rumour for a thread.

Figure \ref{fig:replies-retweets-sandy} shows the interface of the tool we developed for the annotation of social media rumours, where the timeline of tweets is visualised, along with the options to annotate a tweet and visualise the associated conversation.

\begin{figure}[hbt]
 \centering
 \includegraphics[width=0.5\textwidth]{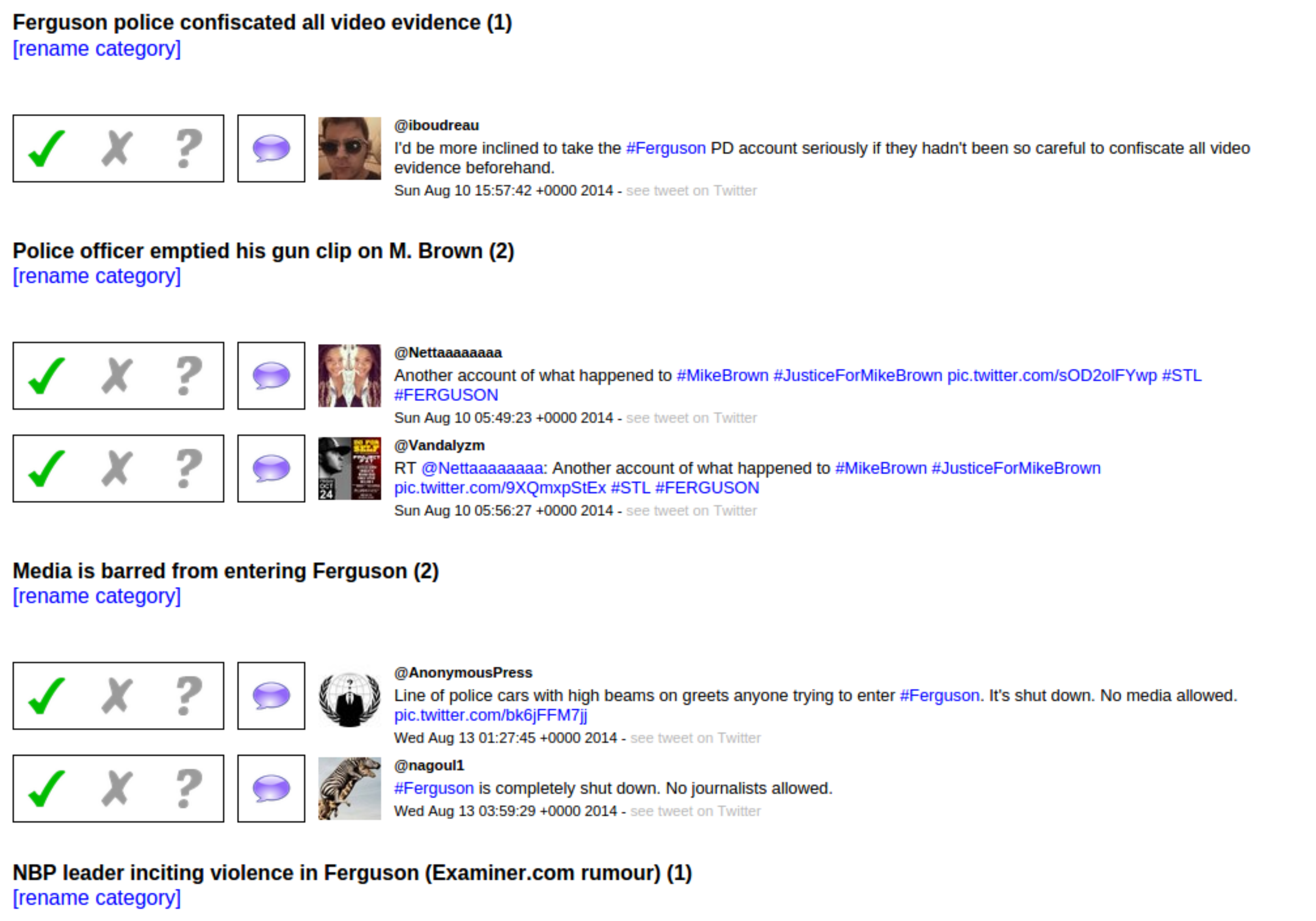}
 \caption{Interface that allows to revise annotations, rename categories, and move threads to another category}
 \label{fig:interface-revise}
\end{figure}

The tool also includes an interface that allows the annotator to review the result of their annotation. The interface summarises the threads annotated as rumours, as well as the stories they were assigned to, which provides a visual summary of what is annotated as a rumour. This is illustrated in Figure \ref{fig:interface-revise}. The interface also makes it easy to rename categories and to move threads to a different category.

\section{Data Collection}

We use Twitter's streaming API to collect tweets, using a set of keywords to filter tweets related to a certain ongoing event. We did this during the Ferguson unrest, which took place in Missouri, USA, after a man named Michael Brown was fatally shot by the police. Rumours emerged in social media as people started protesting in the streets of Ferguson. The event was massively discussed in subsequent days and reported by many different sources in social media. For this event, we tracked the keyword \#ferguson from 9th-25th August 2014, which led to the collection of more than 8.7 million tweets. The hashtag \#ferguson was selected for data collection as the most widely spread hashtag referring to the event\footnote{http://blogs.wsj.com/dispatch/2014/08/18/how-ferguson-has-unfolded-on-twitter/}. In the future we plan to use more sophisticated techniques for adaptive hashtag identification such as \cite{wang2015adaptive}, to retrieve a broader dataset.

\begin{figure}[hbt]
 \centering
 \includegraphics[width=0.25\textwidth]{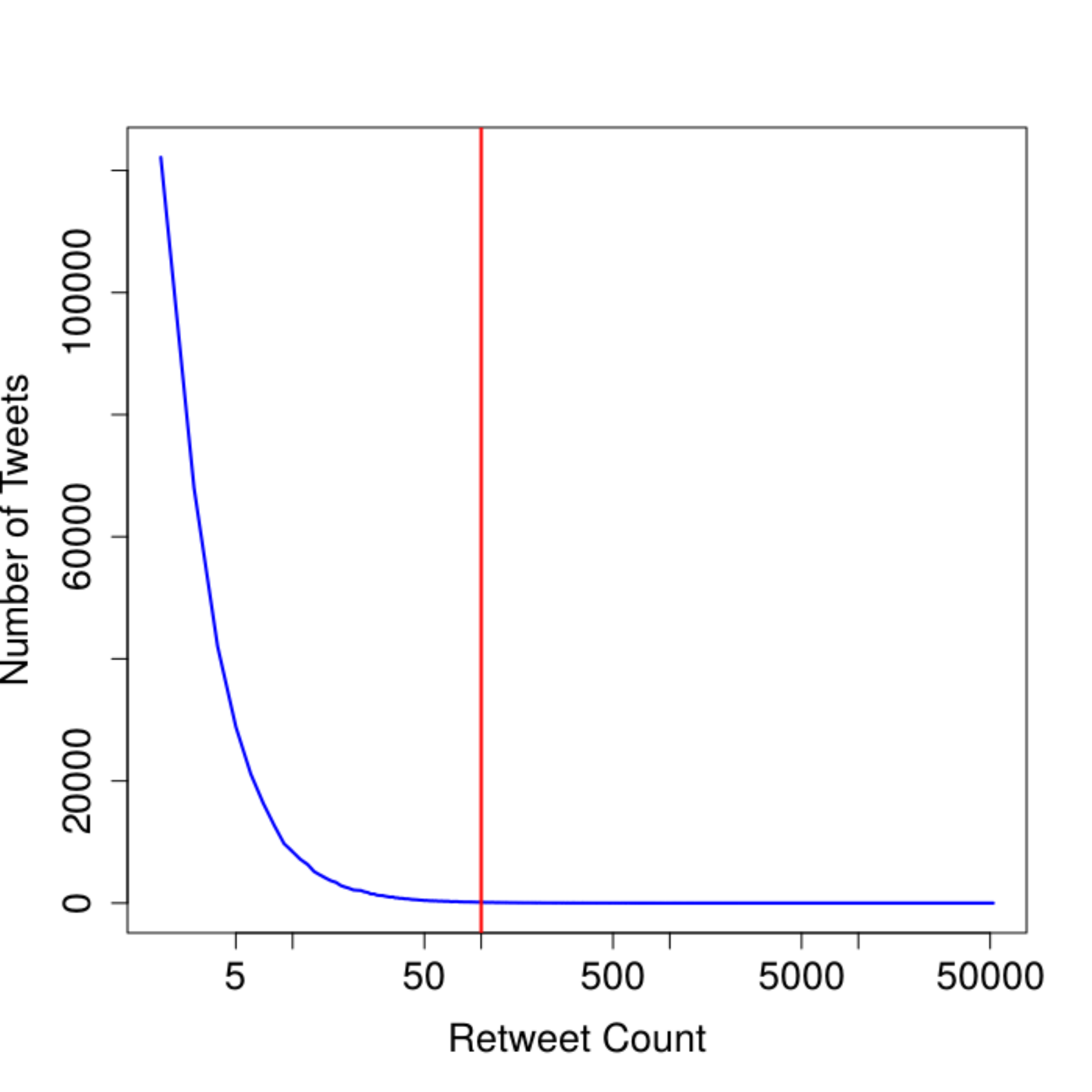}
 \caption{Distribution of retweet counts in the dataset}
 \label{fig:rt-distribution}
\end{figure}

Given the size of this collection of tweets, we filtered it by selecting those tweets that sparked a significant number of retweets, in line with the definition of rumours described above. The threshold for the number of retweets was identified through distributional analysis of retweet counts per tweet (see Figure \ref{fig:rt-distribution}) and empirical tests with different thresholds and was set to 100 retweets for this case. This process makes the dataset more tractable for manual annotation by removing tweets that did not attract significant interest. Finally, to facilitate the annotation task by a native English speaker, we removed non-English tweets. Sampling the Ferguson tweets following these criteria led to a smaller subset of 12,595 tweets. We refer to these as source tweets.

We completed this subset of source tweets by collecting the threads associated with them, i.e., the sets of tweets replying to each of the source tweets. Twitter used to have an API endpoint (called \textit{related\_results}) that allowed the collection of conversations, but this endpoint is no longer available. Thus, we collected conversations by scraping the web page of each of the tweets. This was done iteratively to find even deeper levels of replies (i.e., scraping also the web pages of these replies, to collect replies to those), which enabled us to retrieve the IDs of replying tweets, and then to collect the content of the tweets through Twitter's API. We collected 262,495 replying tweets this way, an average of 20.8 per source tweet. This allows us to visualise the conversations indented by levels as is the case in online fora.

\section{Results}

The annotation task was performed by a team of journalists, SwissInfo, who are members of one of the PHEME project’s partner organisations. To maximise the quality of the annotation, they had discussions within the team during the task. They were instructed to annotate the tweets as being rumours or not by relying on the aforementioned definition of a rumour. The tweets were organised by day, so that clicking on a particular date enabled them to see a timeline of tweets posted that day.

The annotation has been performed for four different days in the dataset: 9th, 10th, 13th and 15th of August. These dates were picked as being eventful after profiling the whole dataset day-by-day. This set of annotations amounts to 1,185 source tweets and threads. The task of annotating these 1,185 conversations took nearly 8 hours in total. From the timestamps we saved with each annotation, we computed the average time needed per thread by removing outliers in the top and bottom 5\% percentiles. The annotation took an average of 23.5 seconds per thread, with an average of 20.7 seconds for those deemed non-rumours, and 31.8 seconds for those deemed rumours. The rumours took longer to annotate than non-rumours not only because they need a second step of assigning to a story, but also because they may also require additional time for the annotator to research the story (e.g. by searching for it on the Web).

The annotation resulted in 291 threads (24.6\%) being annotated as rumours. The distribution of rumours and non-rumours varies significantly across days, as shown in Table \ref{tab:distribution-rumours}. The number of rumours was relatively small in the first few days, always below 15\%, but increased significantly on 15th August, with as many as 45\% rumours. Nevertheless, the number of stories that the tweets were associated with is very similar for the 10th, 13th, and 15th, showing that the number of rumourous threads increased dramatically on the 15th, while the number of rumourous stories remained constant\footnote{Note that the total number of stories, 42, does not match the sum of stories in each of the four days, given that some stories were discussed for more than one day, so we count them only once.}. We believe that the main reason that the number of rumourous tweets surged on the 15th is the emergence of the following three rumours that sparked substantial discussion and uncertainty: (i) that the name of the policeman who killed Michael Brown was about to be announced, (ii) conjecturing about possible reasons why the police may have fatally shot Michael Brown, including that he may have been involved in robbery, and (iii) claims that a new shooting may have taken place in Ferguson, killing a woman in this case. When we look at the distribution of rumours and non-rumours for different numbers of retweets, we observe that the percentage of rumours decreases slightly for tweets with smaller numbers of retweets (i.e., 27.42\% of tweets with at least 250 retweets are rumours, while 24.6\% of tweets with 100 or more retweets are rumours). This decreasing trend suggests that the selection of a threshold is suitable for the annotation of rumours. We also believe that 100 is a suitable threshold for this event, although further looking at lower threshold values would help buttress its validity, which we could not test in this case given the popularity of the event and large scale of the dataset.

\begin{table}[htb]
 \centering
 \begin{tabular}{| l | r | r | r | r | r |}
  \hline
  \textbf{} & \multicolumn{2}{| c |}{Threads} & \multicolumn{2}{| c |}{Thread Sizes} & \\
  \hline
  \textbf{Day} & \textbf{Rum. (\%)} & \textbf{All} & \textbf{Avg.} & \textbf{Med.} & \textbf{Stories} \\
  \hline
  \hline
  9 Aug & 2 (14.3\%) & 14 & 31 & 42 & 2 \\
  \hline
  10 Aug & 18 (8.7\%) & 206 & 16.5 & 16 & 13 \\
  \hline
  13 Aug & 30 (7.0\%) & 430 & 16.3 & 15 & 17 \\
  \hline
  15 Aug & 241 (45.0\%) & 535 & 20.5 & 16 & 17 \\
  \hline
  \hline
  Overall & 291 (24.6\%) & 1,185 & 19.9 & 16 & 42 \\
  \hline
 \end{tabular}
 \caption{Distribution of rumours and stories across days}
 \label{tab:distribution-rumours}
\end{table}

\begin{figure*}[tbh]
  \centering
  \includegraphics[width=0.70\textwidth]{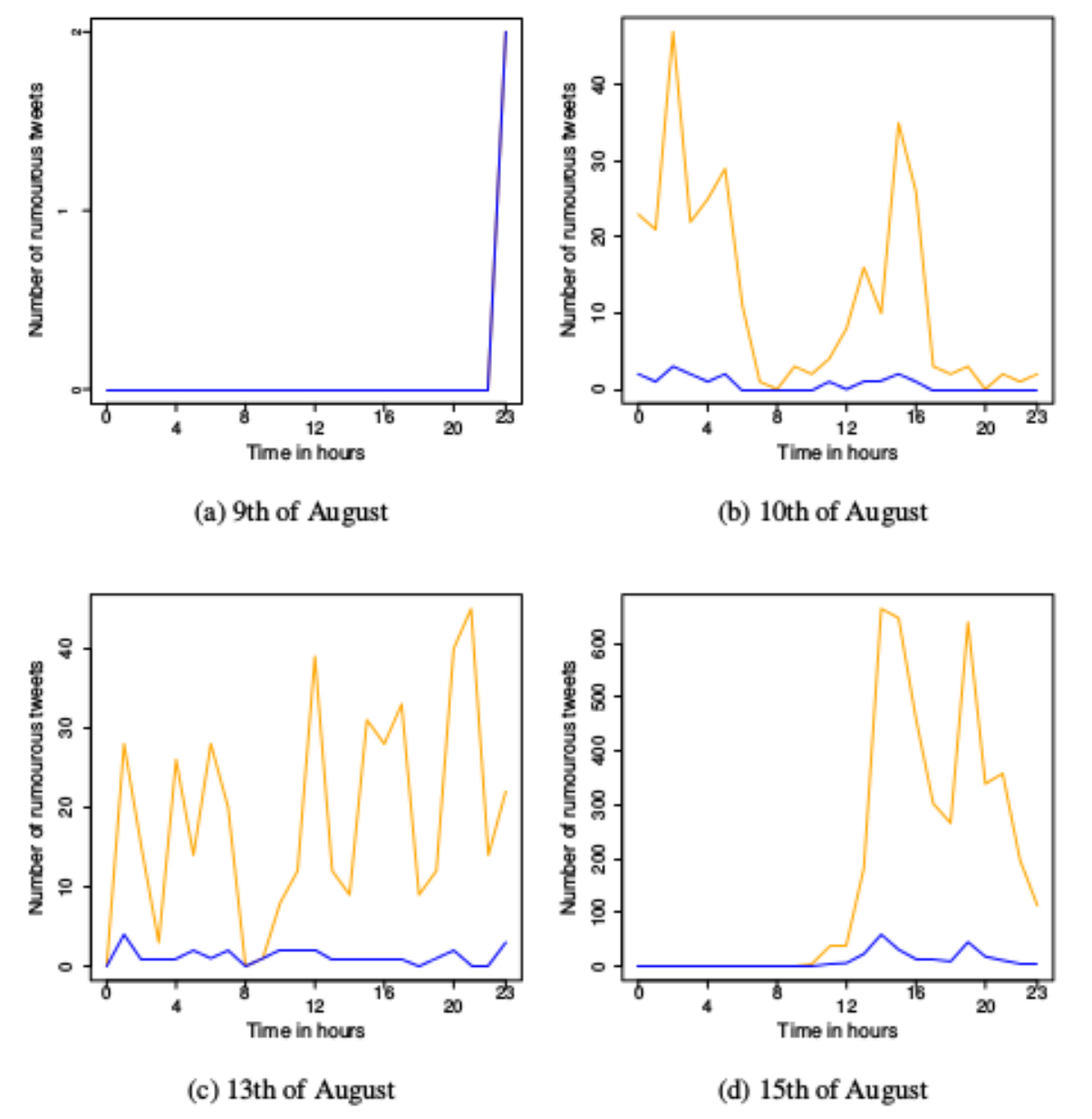}
  \caption{Rumourous source tweets (blue) and replies (orange) across time, with an hour as step size}
  \label{fig:rumours-in-time}
\end{figure*}

Examining the rumourous threads for these days in more detail, Figure \ref{fig:rumours-in-time} shows histograms of their distributions across time for the four days under study. These histograms show very different trends for these days. While rumours were quite uniformly distributed on the 13th of August, there were almost no rumours in the first part of the 15th, with a huge spike of rumours emerging in the afternoon.

\begin{figure}[hbt]
 \centering
 \includegraphics[width=0.35\textwidth]{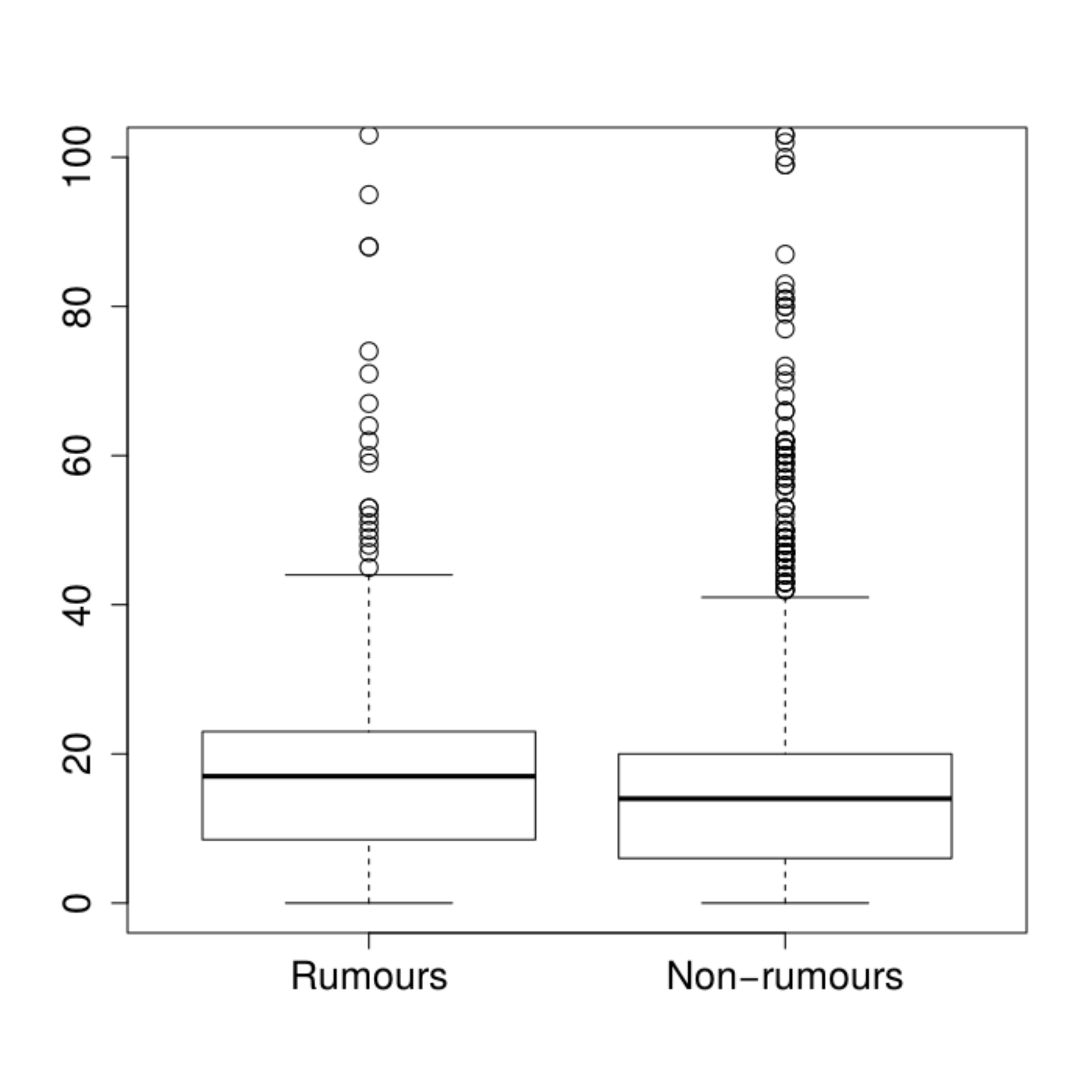}
 \caption{Distribution of conversation sizes for rumours and non-rumours}
 \label{fig:conversation-sizes}
\end{figure}

The 291 source tweets ultimately identified as rumours were categorised into 42 different stories. These stories range from very popular and highly discussed stories such as \textit{Michael Brown having been involved in a robbery} (with 89 threads) or the \textit{potential announcement of the police officer involved in the shooting} (with 26 threads), to lesser discussed stories such as the \textit{Pentagon having supplied St. Louis county police with military-grade weapons} (with 1 thread) or the fact that \textit{two of the four police departments in Ferguson were trained by Israel} (with 1 thread). The fact that we performed the annotation by reading through the timeline of tweets has helped identify not only threads, but also a diverse set of stories that would have been lost if annotation had been driven by a set of manually predefined stories, especially the not-so-popular stories of which we were unaware. This enriches the annotated dataset by broadening the set of rumours.

Having collected the conversations for each of the source tweets annotated, we also compared the extent to which rumours and non-rumours differ in the degree of discussion sparked, by looking at the number of replies they received. We might expect rumours to result in more responses due to being more controversial at the time of posting. Figure \ref{fig:conversation-sizes} shows the distribution of the number of replies in the conversations for rumours and non-rumours. This distribution shows that rumours do provoke slightly more replies than non-rumours, with a slightly higher median. However, non-rumours often generate as many replies as rumours, potentially because of the emotional responses that factual verified information can produce.

\section{Discussion and Future Work}

We have introduced a new definition for rumours and a new method to collect, sample and annotate tweets associated with an event. To implement the method, we have developed an annotation tool. This has allowed us to generate an initial dataset of rumours and non-rumours, which we plan to expand with data from future events. In contrast to related work that predefines a set of rumours and then looks for tweets associated with these, our methodology involves reading through the timeline of tweets to pick out the ones that include rumours and categorise them into stories. This has proven effective for identifying not only a large number of rumourous tweets, but also a diverse set of stories. By looking at 1,185 tweets about the Ferguson unrest in 2014, we have found that 24.6\% were actually rumourous and that these can be categorised into 42 different stories. We aim to expand the dataset and come up with a reasonably large set of rumours, as well as non-rumours.

We believe that the creation of such an annotated dataset of rumours will help to develop tools that make use of machine learning methods to identify rumourous information posted in social media. The automated identification of rumours in social media can in turn be used to help alleviate the spread of misinformation surrounding a situation, which is instrumental in ensuring the well-being of citizens affected by the matter in question. Examples of emergency situations in which the early identification of rumours posted in social media can assist citizens to stay safe include, among others, a natural disaster, a terrorist attack, or riots. Another interesting context for the application of such a rumour detection tool can be during public health emergencies, where the spread of accurate information can be key to calm a worried public \cite{hyer2005effective}. In these situations, citizens need to stay abreast of the latest events to make sure where and how to stay safe in the city, as well as to know the state of certain services such as public transportation. Similarly, reducing the spread of misinformation and emphasising accurate information can be extremely useful not only to journalists and others who need to keep citizens informed, but also government staff who need to take the right decisions at the right time to maximise safety within a city.

Having collected threads sparked by each of the source tweets manually annotated as rumours and non-rumours, we are developing an annotation scheme to help determine the contribution of each of the tweets in the thread/conversation to the story. This will also allow us to study the effectiveness of Twitter's self-correcting mechanism, among others, by looking at the evolution of a rumour within associated conversations. To do so, the annotation scheme will look at how each of the tweets supports or denies a rumour, the confidence of the author, as well as the evidence provided to back up their statements. The creation of such datasets with annotated conversations will then enable us to develop machine learning and natural language processing tools to deal with misinformation in these situations.

\section{Acknowledgments}

We would like to thank the team of journalists at SwissInfo for the annotation work. This work has been supported by the PHEME FP7 project (grant No. 611233).

\bibliographystyle{aaai}
\bibliography{aicities-rumours}

\end{document}